# The Highest Melting Point Material: Searched by Bayesian Global Optimization with Deep Potential Molecular Dynamics

Fu-Zhi Dai [1,2,*], Bo Wen [3], Xingjian Jiao [3], Lei Chen [4,5], Yujin Wang [4,5,*]

[1] *Artificial Intelligence for Science Institute, Beijing 100080, China*

[2] *DP Technology, Beijing, 100080, China*

[3] *Science and Technology on Advanced Functional Composite Laboratory, Aerospace Research Institute of Materials & Processing Technology, Beijing 100076, China*

[4] *Institute for Advanced Ceramics, School of Materials Science and Engineering, Harbin Institute of Technology, Harbin 150001, China*

[5] *Key Laboratory of Advanced Structure-Function Integrated Materials and Green Manufacturing Technology, Harbin Institute of Technology, Harbin 150001, China*

* Corresponding author.

*E-mail address:* dfzshiwo@163.com (Fu-Zhi Dai), wangyuj@hit.edu.cn (Yujin Wang).

The interest in refractory materials is increasing rapidly in recent decades due to the development of hypersonic vehicles. However, which substance has the highest melting point keeps a secret, since precise measurements in extreme condition are overwhelmingly difficult. In the present work, an accurate deep potential model of Hf-Ta-C-N system was firstly trained, and then applied to search for the highest melting point material by using molecular dynamics simulation and Bayesian global optimization. The predicted melting points agree well with experiments, and confirm that the carbon site vacancy can enhance melting points of rock-salt structure carbides. Solid solution with N is verified as another new and more effective melting point enhancing approach for HfC, while the conventional routing of solid solution with Ta (*e.g.* HfTa$_4$C$_5$) is not suggested to result in a maximum melting point. The highest melting point (~ 4236 K) is achieved with composition of HfC$_{0.638}$N$_{0.271}$, which is ~ 80 K higher than the highest value in Hf-C binary system. The dominating mechanism of N addition is believed to be the instable C-N and N-N bonds in liquid phase, which reduces the liquid phase entropy and renders the liquid phase less stable. The improved melting point and fewer gas generation during oxidation by addition of N provides new routing to modify the thermal protection materials for hypersonic vehicles.

# 1. Introduction

The interest in refractory materials is increasing rapidly in recent decades due to the development of new generation of aerospace vehicles, where high temperature materials are widely applied in thermal protection systems, propulsion system, and *etc*. [1-7]. However, which substance has the highest melting point keeps a secret, the answer of which is of both scientific interest and technical demanding. It is well-known that tungsten has the highest melting point (~ 3700 K) among all the elements at ambient pressure, which guarantees its widespread usage in high temperature techniques. Another well-known refractory element is carbon, which sublimates instead of melting at ambient pressure. It means that the melting point of carbon is not definable at ambient pressure. Thus, there is no doubt on which element has the highest melting point. However, instead of the highest melting point element, we concern about which substance has the highest melting point? It is a controversial question for a long history, which has not reached an agreement until now.

People may blurt out the phrase "$HfTa_4C_5$ has the highest melting point of any known material: 4215 °C", which has become ingrained in countless textbooks, and even in the 14th edition of Encyclopaedia Britannica [8-10]. However, the phrase results from a mistake due to unit conversion error, where the melting point of $HfTa_4C_5$ was reported to be "4215 in °abs" in the original paper of Agte and Alterthum, which is 4215 K [10,11]. Similar result that the solid solution between tantalum and hafnium carbide with Ta:Hf ~ 4:1 exhibits the maximum [12] or a local maximum melting point [13] were also reported by many other experiments. Different from this viewpoint, there is another voice, which suggests that either tantalum or hafnium carbide has the highest melting point instead of their solid solution. In 1960s,

Rudy *et al.* [14-16] carried out systematic investigations on high temperature materials, which were motivated by the development of hypersonic vehicles. The phase diagrams and data from the project leaded by Rudy are still the most used by researchers working in these materials. They reported that the highest melting point of tantalum and hafnium carbides appear at nonstoichiometric compositions [14-16]. $TaC_{0.88}$ has the highest melting point 4256±15 K in tantalum carbides [14], while $HfC_{0.94}$ has the highest melting point 4201±20 K in hafnium carbides [15], which was later on corrected to 4223±20 K to account for zirconium impurities in the original hafnium used to prepare the carbides [16]. In addition, they revealed that melting point decreased monotonously from $TaC_{0.88}$ to $HfC_{0.94}$ without any local maximum in between. Even though the topic has been discussed for almost 100 years since the first report by Agte and Alterthum in 1930, experimental measurements can hardly achieve agreement due to the following difficulties:

(1) The real composition of a sample cannot be accurately controlled. On the one hand, there are many impurities in the sample, and the effects of the impurities on the melting point are unknown. On the other hand, carbon prefers to evaporate during heating, resulting in a different composition when reaching the melting point. As a result, the correlation between the melting point and sample composition is not accurate.

(2) Accurate temperature measurement at ultra-high temperatures is difficult. For example, corrections are necessary when converting observed temperatures to true temperatures, or true temperatures are derived based on the emissivity of a material, the exact value of which is not accurately known. Usually, emissivity at other temperatures instead of that at the melting point is used [13, 17]. Therefore, there are evident uncertainties in experimentally reported melting points.

Theoretical investigation is a good complementary to experiments, which does not have the above limitations of experiments. Hong *et al.* [18] adopted *ab initio* molecular dynamics (AIMD) to evaluate melting points of tantalum and hafnium carbides and carbonitrides. They revealed that the calculated melting temperature dependence on carbon content agree with experiments, indicating the capability of theoretical approaches. Moreover, they reported that addition of nitrogen may increase the melting temperature of hafnium carbides. However, due to the limited simulation size (no more than 64 atoms in the simulations by Hong *et al.* [18]) and time in AIMD, the composition and transition probability may not be adequately sampled. To overcome the shortcomings in AIMD, we applied Deep Potential molecular dynamics (DPMD) and Bayesian global optimization to search for the highest melting point substance. In section II, the generation process and validation of the DP model are introduced. In section III, the dependence of melting point on composition in rock-salt Hf-Ta-C-N system are determined and the highest melting point composition is searched by using Bayesian global optimization.

## 2. Generation and validation of the DP model

Even though it is still a controversy about what substance has the highest melting point, there is no doubt that the highest melting substance is amongst tantalum and hafnium carbides or carbonitrides. Therefore, we focused on evaluation the composition dependent melting point in rock-salt structure Hf-Ta-C-N system. To adequately sample the composition, we need molecular dynamics (MD) simulation methods with both high accuracy and high efficiency. Thanks to the developments in artificial intelligent, machine learning potentials [19-23] bridge the gap between density functional theory (DFT) based methods and MD simulations, which guarantee atomistic simulations with accuracy comparable to DFT based methods and at cost comparable to MD simulations.

*2.1 Generation of the DP Model*

The machine learning potential proposed by Zhang *et al.* [23,24], named Deep Potential (DP), was adopted to train an interatomic potential for Hf-Ta-C-N system from a dataset generated by DFT calculations. The dataset was explored by a concurrent learning scheme that is implemented in the DP-Generator (DP-GEN) software [25,26]. The DP-GEN software explores the configurational space, including both element distributions and conformation arrangements, iteratively by three steps: training, exploration, and labeling. During training, four different DP models with different activation functions and initialization parameters were trained based on existing data. Then, hybrid MD and Monte Carlo (MC) simulations under isothermal-isobaric (NPT) ensemble were performed to sample the configurational space. The explored thermodynamic states span the temperature range [100, 6000] K and the pressure range [-5, 10] GPa. When T > 3000 K, only positive pressure was applied.

The prediction accuracy of a configuration is measured by "model deviation", which is defined as the maximal standard deviation of forces predicted by the four DP models. Candidate configurations were randomly chosen from the simulation trajectories if their model deviations were in a predefined range [$\varepsilon_{low}$, $\varepsilon_{high}$]. [$\varepsilon_{low}$, $\varepsilon_{high}$] was [0.3, 0.5] when T < 3000 K, and [0.5, 1.0] when T > 3000 K. The selected configurations were then calculated by using Vienna ab initio simulation package (VASP) [27,28]. The exchange-correlation energy was modeled by the Perdew-Burke-Ernzerhof (PBE) functional [29]. The projector-augmented-wave (PAW) method [30,31] is used. The kinetic energy cut-off of the plane wave was set to be 900 eV. $k$-points mesh according to Monkhorst-Pack method [32] with a separation of 0.15 Å$^{-1}$ was adopted in the Brillouin zone. The self-consistent field iteration stops when the difference in total energy of consecutive iterations is less than $10^{-6}$ eV. The DP-GEN iteration process was stop when the prediction accuracy of the DP model was higher than 97% at each thermodynamic condition.

After collecting the dataset, the DP model was trained by using the DeePMD-kit software [24]. The DP model maps the local atomic configurations to atomic energies by deep neural networks [23,24], and has been proven to appliable to many materials [33-39]. The DP model contains two sets of neural networks. The first one is a descriptor net, which automatically encodes local atomic configurations to symmetry-preserving descriptors. The second one is a fitting net, which maps the symmetry-preserving descriptors to atomic energies. The architecture of the DP model is set as follows:

(1) The smooth edition descriptor from reference [23] is used, which consists of 3 layers of neural networks with each layer having 25, 50, and 100 nodes. The projection dimension is set to 12. The cutoff is set to be 7.0 Å with a smooth function imposed from 2.0 Å. The "type-one" architecture is adopted to reduce the scale of the descriptor net.

(2) The fitting net includes 3 layers of neural network with each layer having 240, 240, and 240 nodes.

The model was trained by two steps. Firstly, the model was trained from scratch. The learning rate decays from $1.0 \times 10^{-3}$ to $1.0 \times 10^{-8}$, and the per-factors of energy, force and virial in loss functions change from 0.01 to 1, from 100 to 1, and from 1 to 1, respectively. Then, the model was re-trained. In this step, the learning rate decays from $1.0 \times 10^{-4}$ to $1.0 \times 10^{-8}$, and the per-factors of energy, force and virial in loss functions are set to be 10, 1, and 1 without any change, respectively. For details of the parameters, please refer to the original papers and the open-source software [23,24].

## 2.2 Validation of the DP Model

The accuracy of the DP model was checked by comparing with DFT calculations. Fig. 1(a) compares the energies predicted by the DP model with DFT calculations, and Fig. 1(b) illustrates the distribution of prediction error in energy. Fig. 1(c) compares the predicted forces by the DP model with DFT calculations, and Fig. 1(d) shows the error distribution of prediction error in force. The results reveal that the DP model agrees well with DFT calculations, where the prediction error in energy and force are 3.9 meV/atom and 172 meV/Å, respectively. The prediction errors are comparable to many other similar systems as reported [36-38]. Table 1 compares the lattice parameter (*a*) and elastic constants ($C_{11}$, $C_{12}$, and $C_{44}$) of rock-salt structure HfC, TaC, HfN and TaN predicted by the DP model and the DFT calculations, where excellent agreements can also be found. Fig. 2 compares the equation of state of rock-salt structure HfC, HfN, TaC and TaN precited by the DP model and DFT calculations, which all show good agreements. These comparisons reveal the accuracy of the DP model, which guarantee the trustable predictions in following sections.

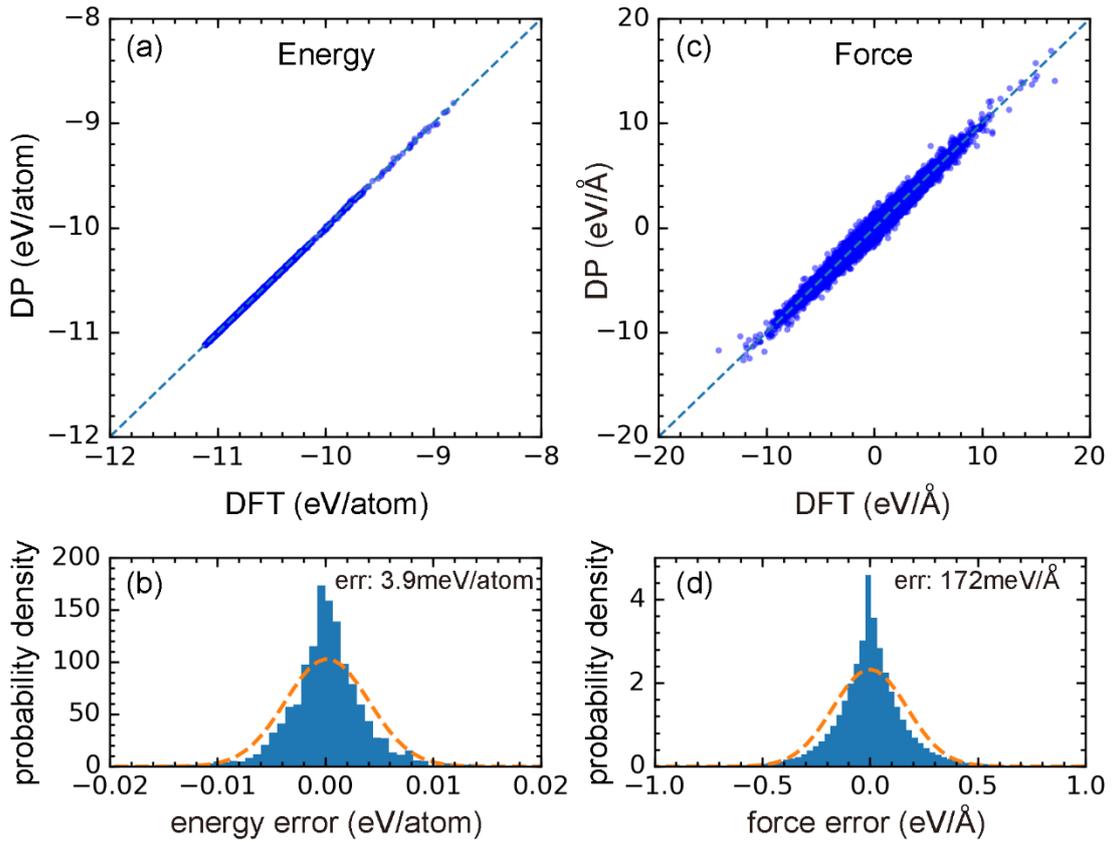

Figure 1. Comparison of (a) energies and (c) forces predicted by the DP model with DFT calculations, and probability density distributions of prediction errors on (b) energy and (d) force.

Table 1. Comparison of DP prediction results with DFT calculations. The DFT values in this table were calculated with the spacing of $k$-point mesh being 0.05 Å$^{-1}$.

|  |  | $a$ (Å) | Energy (eV/atom) | $C_{11}$ (GPa) | $C_{12}$ (GPa) | $C_{44}$ (GPa) |
|---|---|---|---|---|---|---|
| HfC | DFT | 4.647 | -10.5253 | 513 | 105 | 172 |
| HfC | DP | 4.647 | -10.5255 | 513 | 97 | 162 |
| TaC | DFT | 4.479 | -11.1039 | 708 | 134 | 176 |
| TaC | DP | 4.478 | -11.1032 | 704 | 116 | 166 |
| HfN | DFT | 4.535 | -10.8823 | 593 | 109 | 119 |
| HfN | DP | 4.536 | -10.8824 | 587 | 116 | 113 |
| TaN | DFT | 4.421 | -10.9177 | 718 | 137 | 60 |
| TaN | DP | 4.420 | -10.9169 | 678 | 141 | 55 |

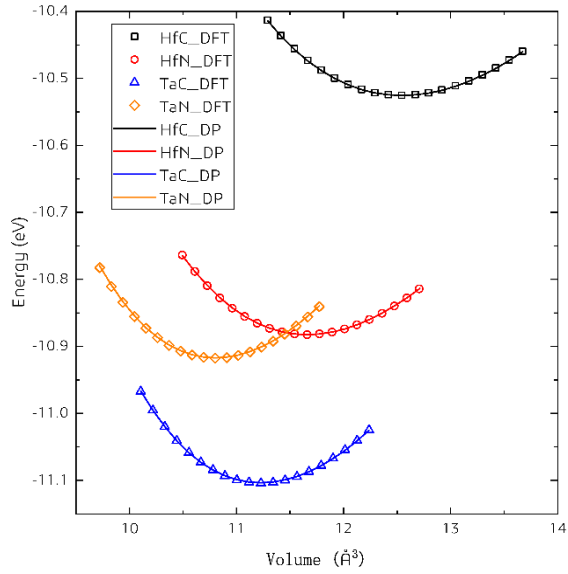

Figure 2. Comparison of the equation of state of rock-salt structure HfC, HfN, TaC and TaN precited by the DP model and DFT calculations.

## 3. Results and Discussions

Solid-liquid coexistence MD simulation was adopted to evaluate the melting point of a given composition. The size of the simulation box was 5×5×10 (X×Y×Z) unitcells, and the interface between the solid and liquid phase was along the $Z$ direction. Timestep in MD simulations was set to be 1 fs. Isothermal-isobaric (NPT) ensemble was adopted during simulations. The system was firstly equilibrated at 3500 K for 50 ps, and half of the system was heated to 5000 K to melt and equilibrated for 50 ps with the other half frozen. Then, the whole system was set to different temperatures and equilibrated for 100 ps to search for the melting point. MD simulations were implemented with LAMMPS software [40,41] compiled with DeePMD-kit package. Bayesian global optimization [42] was used to efficiently search for the composition with the highest melting point.

*3.1 Melting points in $HfC_{1-x}$ and $TaC_{1-x}$*

In HfC or TaC, it has reached a consensus that the highest melting point appears at a nonstoichiometric composition with the depletion of C. Here, we checked the reliability of the DP model in prediction this phenomenon, and the accuracy in predicting melting points. The melting point of a series of $HfC_{1-x}$ and $TaC_{1-x}$ were calculated, where $x$ changes from 0 to 0.2 with a step of 0.02. The predicted results are shown in Fig. 3, together with experimental measurements [14-16,43] and results calculated by Hong *et al.* using AIMD [18] for comparison. It is clear that the predicted results by the DP model agree well with experiments, which confirm that nonstoichiometric composition with the depletion of C has the highest melting point in both $HfC_{1-x}$ and $TaC_{1-x}$. For $HfC_{1-x}$, the maximum melting point is reached at $x \sim$ 0.12. For $TaC_{1-x}$, the highest melting point composition is $x \sim 0.14$. The highest melting point compositions in experiments by Rudy *et al.* [14-16] are $x \sim 0.06$ for

$HfC_{1-x}$ and $x \sim 0.12$ for $TaC_{1-x}$. The predicted highest melting point of $HfC_{1-x}$ is $\sim 4160$ K, $\sim 60$ K lower than the experimental result reported by Rudy *et al.* [14-16], while the predicted highest melting point of $TaC_{1-x}$ is $\sim 4050$ K, $\sim 200$ K lower than the experimental result reported by Rudy *et al.* [14-16]. In comparison to results calculated by Hong *et al.* using AIMD [18], our calculated melting points are $\sim 200$ K higher, which agree much better with experiments. In addition, the highest melting compositions predicted by DPMD match better with experiments by Rudy *et al.* [14-16] in comparison to compositions predicted by AIMD by Hong *et al.* [18]. For example, the highest composition in $HfC_{1-x}$ is $x \sim 0.12$ in DPMD, $x \sim 0.06$ in experiments by Rudy *et al.* [14-16], and $x \sim 0.19$ in AIMD by Hong *et al.* [18], as shown in Fig. 3(a). The better agreements of our simulations may result from the more adequate sampling with bigger simulation system and longer simulation time. In calculations, both our results and results reported by Hong *et al.* [18], the melting points of $HfC_{1-x}$ are $\sim 100$ K higher than those of $TaC_{1-x}$. In experiments, it has been reported that either $TaC_{1-x}$ or $HfC_{1-x}$ had slightly higher melting point. For example, Rudy *et al.* [14-16] reported that the melting points of $TaC_{1-x}$ were slightly higher, while Cedillos-Barraza *et al.* [13] reported that the melting points of $HfC_{1-x}$ were slightly higher. Due to the errors in DFT calculations, the DP model and experimental measurements, it can hardly say whether the melting points of $HfC_{1-x}$ or $TaC_{1-x}$ are higher.

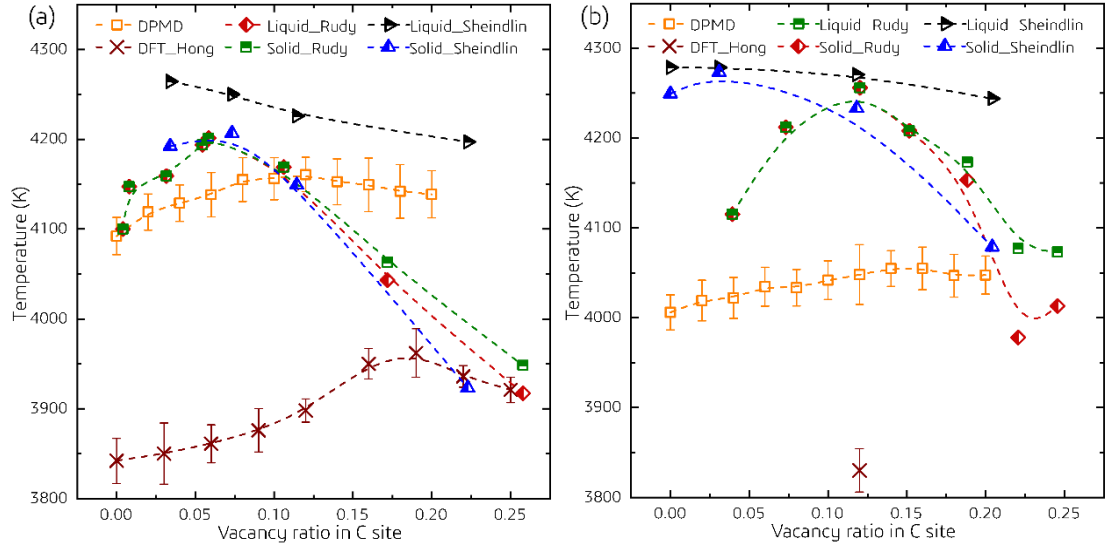

Figure 3. Comparison of calculated and measured melting points of (a) $HfC_{1-x}$ and (b) $TaC_{1-x}$. DPMD represents results calculated in this work. DFT_Hong means results calculated by Hong *et al.* using DFT [18]. Liquid_Rudy and Solid_Rudy [14-16], Liquid_Sheindlin and Solid_Sheindlin [42], are experimental liquid and solid lines measured by the corresponding authors.

*3.2 Melting points in $Hf_{1-y}Ta_yC_{0.88}$*

It has long been a controversy that whether there is a local or global maximum melting point in solid solutions between $HfC_{1-x}$ and $TaC_{1-x}$. To clarify this controversy, the melting point of a series of $Hf_{1-y}Ta_yC_{0.88}$ were calculated, where $y$ changes from 0.0 to 1.0 with a step of 0.1. As shown above in Fig. 4, both $HfC_{1-x}$ and $TaC_{1-x}$ roughly have their highest melting point with $x \sim 0.12$. Therefore, the vacancy concentration in C site was set to be 12%. The predicted results are shown in Fig. 4, together with experimental results [11-16] and results calculated by Hong *et al.* using AIMD [18] for comparison. Our results reveal that the melting point changes monotonously between HfC$_{0.88}$ and TaC$_{0.88}$ without any local maximum or minimum in between. However, the variation of melting point with respect to Ta content predicted by Hong *et al.* [18] is rugged, which may result from the inadequate sampling by AIMD. Thus, it can hardly say whether there is local or global maximum or minimum in between or not. Our simulation results agree with results reported by Rudy *et al.* [18], which do not support the conclusion that HfTa$_4$C$_5$ exhibits the highest melting point, even though other results argued that local or global maximum melting point existed in HfC$_{1-x}$ and TaC$_{1-x}$ solid solutions.

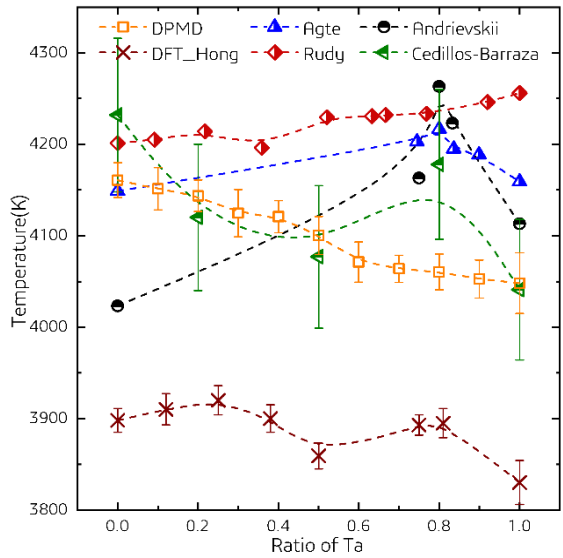

Figure 4. Comparison of calculated and measured melting points of Hf$_{1-y}$Ta$_y$C$_{1-x}$. DPMD represents results of Hf$_{1-y}$Ta$_y$C$_{0.88}$ calculated in this work. DFT_Hong means results calculated by Hong *et al.* using DFT [18]. Agte [11], Rudy [14-16], Andrievskii [12], and Cedillos-Barraza [13] are experimental results by the corresponding authors.

*3.3 Searching for the highest melting point by Bayesian global optimization*

Hong *et al.* [18] has reported that addition of N may increase the melting point of $HfC_{1-x}$. To search for the possible highest melting point composition, we calculated the melting point of $Hf_{1-y}Ta_yC_{1-x-z}N_z$ with $y$ in [0, 1], $x$ in [0, 0.5] and $z$ in [0, 0.3], where $x$ is the vacancy concentration in C site. The smallest step in $x$, $y$ and $z$ are all 1/500 according to the simulation cell size, resulting in a quite large searching space with $10^6$ in magnitude. Bayesian global optimization (BGO) was adopted to improve the searching efficiency, the process of which is illustrated in Fig. 5(a). The software package "BayesianOptimization" [42] was adopted to implement BGO. BGO searches for the highest melting point in an iterative way. Firstly, a surrogate model that fits the dependence of melting point $T_m$ on the composition ($x$, $y$, $z$) based on the existing data is trained. Then, a new data point is calculated based on the guidance of the surrogate model and a utility function. Gaussian process was adopted as the surrogate model here. The upper confidence bound with $\kappa = 2.5$ was adopted as the utility function, which means that the next composition to try is the one has the highest $\mu + \kappa\sigma$. $\mu$ and $\sigma$ are the mean and standard error of the fitted Gaussian process model. With the iterative addition of new data points and improvements on the surrogate model, the highest melting point will be found in an efficient way. The search was stopped when no more higher melting point can be found in 50 iterations.

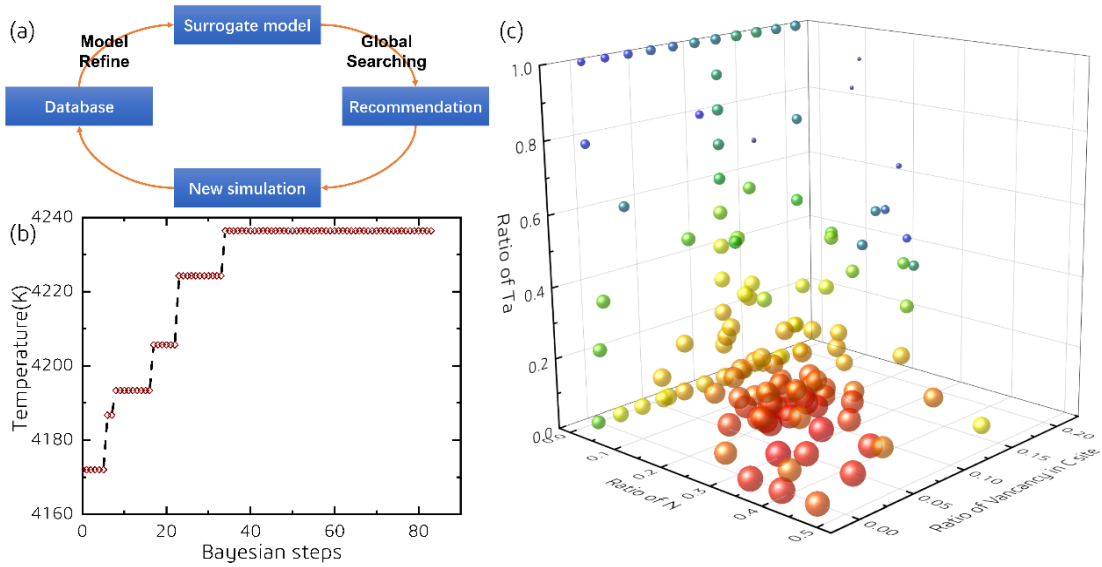

Figure 5. (a) Schematic illustration of Bayesian global optimization. (b) The variation of the searched highest melting point as a function of iteration steps. (c) The variation of melting point with respect to the composition $Hf_{1-y}Ta_yC_{1-x-z}N_z$. In (c), the bigger the sphere is, the higher the melting point is. The smallest melting point and the highest melting point in (c) are 3908 K and 4236 K, respectively.

The variation of the highest melting point in the dataset with respect to the iteration step is shown in Fig. 5(b). It reveals that the BGO can search for the highest melting point in an efficient way, where the highest melting point increases by ~ 60 K within 30 steps. The highest melting point obtained is 4236 K, with its corresponding composition being $x = 0.000$, $y = 0.271$, and $z = 0.091$. Fig. 5(c) shows the relationship between the melting point $T_m$ and the composition ($x$, $y$, $z$). The bigger the point is, the higher the melting point is. In agreement with results reported by Hong *et al.* [18], addition of N can efficiently increase the melting point of $HfC_{1-x}$, even though the melting point of HfN (~ 3705 K by DPMD, and ~ 3663 K in experiments) is much lower than that of HfC. The addition of N plays similar tendency to C site vacancy on the melting point of HfC, where the melting point first increases with and then decreases with the addition of N, resulting in a maximum value. The increasing effect on melting point due to N addition (increased by ~ 120 K at most in comparison to HfC) is more significant than that of introducing C site vacancy (increased by ~ 60 K at most in comparison to HfC). The incorporation of both N addition and C site vacancy results in the highest melting point. Different from HfC, addition of N will reduce the melting point of TaC, as can be seen from the in Fig. 5(c), where TaC$_{1-x}$ are not the compositions with the lowest melting point. The reason for the decreased melting point may be that the rock-salt TaN is not a stable phase. Thus, addition of N in TaC reduces the stability.

*3.4 Discussions*

It is well-known that higher melting point can be achieved by nonstoichiometric composition with depletion of C in HfC and TaC, which is also confirmed by our simulations. However, the longtime controversy that whether there is a highest melting point solid solution between HfC and TaC still need further investigation. Our simulations do not support this assumption, and suggest a monotonous change of melting point between HfC and TaC, which agrees with experimental results of Rudy *et al.* [14-16]. Despite of C site vacancy, we verified another intriguing melting point enhancing mechanism discovered by Hong *et al.* [18]. Substitution C by N is more effective in improving the melting point of HfC in comparison to C site vacancy. As suggested by Hong *et al.* [18], addition of N remarkably changes the liquid structure due to the instability of C-N and N-N bonds. A liquid is stabilized by its higher entropy to offset its higher enthalpy. In particular, the higher entropy of a liquid is reflected by its large variety of pair-wise correlations. For example, the exceeding entropy of a liquid with respect to ideal gas is dominated by the two-body exceeding entropy, which is:

$$S_{ex}/k_B = -2\pi\rho \sum_{i,j} x_i x_j \int_0^\infty \{g_{ij}(r) ln g_{ij}(r) - [g_{ij}(r) - 1]\} r^2 dr$$

where $\rho$ is atomic number density, $x_i$ and $x_j$ are fractional composition, and $g_{ij}$ is pair correlation function between element $i$ and $j$. Fig. 6 shows the pair correlation function of HfC, HfN, and HfC$_{0.5}$N$_{0.5}$ liquids obtained at 4500 K and their two-body exceeding entropy. It is evident that the dominated first neighbor of N is Hf, indicating low stability of C-N and N-N bonds in the liquid phase. In contrast, C-C bond is stable, forming a strong first neighbor correlation peak, where even C-chains with more than three C atoms were found in the simulated atomic structure. The reduced variety of pair-wise correlations with addition of N leads to lower entropy of the liquid phase (*see* Fig. 6(d)), and renders the liquid phase less stable. Moreover, solid solution of N will increase the solid-state configurational entropy, and renders the solid phase more stable. As a result, addition of N enhances the melting point.

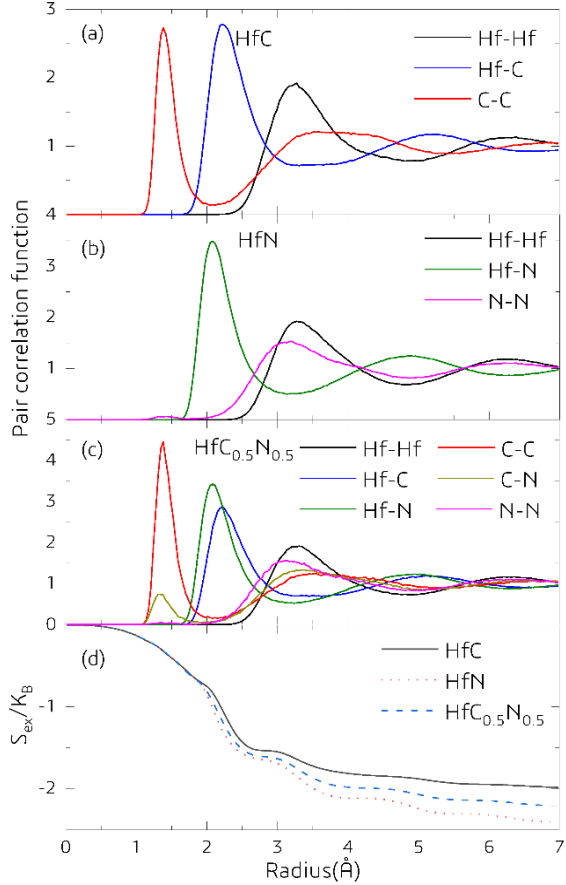

Figure 6. Pair correlation function (normalized as $r \to \infty$) in liquid-state (a) HfC, (b) HfN and (c) $HfC_{0.5}N_{0.5}$. (d) Comparison of the two-body exceeding entropies of liquid-state HfC, HfN and $HfC_{0.5}N_{0.5}$.

Even though almost 100 years have passed since the first report by Agte and Alterthum in 1930 [11], it is still non-trial to answer the question "which substant has the highest melting point at ambient pressure?", since there are large errors in both experiments and simulations. In experiments, the composition of a material cannot be exactly controlled or determined due to the difficulties in sample preparation, sample interaction with crucible and sample evaporation during heating; the determination of real temperature is also problematic due to the extreme conditions and equipment instability, *etc*. For example, Andrievskii *et al.* [12] reported that some of their samples contained high content of N and O, around 10% that of C; Cedillos-Barraza *et al.* [13] also reported similar high impurities in their HfC sample; Savvatimskiy *et al.* [44] detected high content of O in their sample. These impurities may lead to substantial shifts in melting points, and the relation between composition and melting point is not reliable. In early years, the samples were heated by Joule heating, the slow heating effect of which would lead to preferred evaporation of C, resulting in a different composition when approaching the melting point. Laser heating technology developed in recent years can be conducted under containerless condition [13,17,43,44], which overcome the problem of composition shift during heating. However, the narrow time window further challenges the temperature measurement. Cedillos-Barraza *et al.* [13] and Savvatimskiy *et al.* [17] using different emissivity to estimate the temperature, and suggested fully different conclusion on whether HfTa$_4$C$_5$ had the highest melting point. In simulations, there are also many errors, *e.g.* insufficient energy cutoff and *k*-spacing, intrinsic error due to exchange-correlation functionals, fitting error of interatomic potentials, inadequate sampling in simulations, *etc*. Hong *et al.* [18] estimated in their work that the simulated melting points may shift upwards by ~ 460 K when using Heyd-Scuseria-Ernzerhof (HSE) hybrid functional to replace Perdew-Burke-Ernzerhof (PBE) functional. Melting points in our simulations shifted upwards by ~ 200 K in comparison to results calculated by Hong *et al.* [18], which may result from different simulation approaches and the different accuracy of calculation methods. To obtain more reliable predictions, interatomic potentials fitted to more accurate quantum mechanical data, *e.g.* using Heyd-Scuseria-Ernzerhof (HSE) hybrid functional, are needed. In addition, more accurate simulation methods that can obtain thermodynamic equilibrium between solid and liquid phase are also necessary, *e.g.* determining the chemical potential of each elements by simulations. Nevertheless, our simulation results can still provide valuable information on understanding the dependence of melting point on compositions and searching for high melting point materials for applications in extreme conditions, *e.g.* thermal protection system in hypersonic vehicles. HfC is one of the promising compounds applied in thermal protection systems of hypersonic vehicles due to its high melting point and high melting point of its oxidation product, $HfO_2$. The addition of N on one hand will improve the melting point of HfC, and one the other hand will reduce the amount of gases generated during oxidation, resulting in better oxidation scales, and may render better thermal protection effect.

## 4. Conclusions

In the present work, the highest melting point challenge was investigated by molecular dynamics simulations based on an accurate deep potential model of Hf-Ta-C-N system. The predicted melting points are consistent with experimental measurements, indicating the reliability of the simulations. Our results confirmed the well-known phenomenon that C site vacancy is a melting point enhancing mechanism in rock-salt structure transition metal carbides. The longtime controversy that whether the solid solution $HfTa_4C_5$ is the highest melting point substance keeps unsolved, even though our simulations do not support this assumption. More precise methods are need to clarify this assumption in the future. In HfC, addition of N is verified as another melting point enhancing approach, which is more effective than C site vacancy. Addition of N remarkably reduces the pair correlation of C-N and N-N, reduces the entropy of liquid phase, and renders the liquid phase less stable. The enhanced melting point and less gas generation during oxidation by addition of N provides new routing to modify the thermal protection materials for hypersonic vehicles.

## Acknowledgement

This work was supported by Natural Sciences Foundation of China under Grant No. .

## References

[1] W.G. Fahrenholtz, G.E. Hilmas, I.G. Talmy, J.A. Zaykoski, J. Am. Ceram. Soc. 90 (2007) 1347–1364.

[2] Y. Katoh, G. Vasudevamurthy, T. Nozawa, L.L. Snead, J. Nucl. Mater. 441 (2013) 718–742.

[3] M.M. Opeka, I.G. Talmy, J.A. Zaykoski, J. Mater. Sci. 39 (2004) 5887-5904.

[4] H. Xiang, Y. Xing, F.-Z. Dai, H. Wang, L. Su, L. Miao, G. Zhang, Y. Wang, X. Qi, L. Yao, H. Wang, B. Zhao, J. Li, Y. Zhou, J. Adv. Ceram. 10 (2021) 385–441.

[5] J. Binner, M. Porter, B. Baker, J. Zou, V. Venkatachalam, V.R. Diaz, A. D'Angio,

P. Ramanujam, T. Zhang, T.S.R.C. Murthy, Int. Mater. Rev. 65 (2019) 1–56.

[6] D. Ni, Y. Cheng, J. Zhang, J.-X. Liu, J. Zou, B. Chen, H. Wu, H. Li, S. Dong, J. Han, X. Zhang, Q. Fu, G.-J. Zhang, J. Adv. Ceram. 11 (2022) 1–56.

[7] B. Liu, J. Zhao, Y. Liu, J. Xi, Q. Li, H. Xiang, Y. Zhou, J. Mater. Sci. Technol. 88 (2021) 143–157.

[8] J. Emsley, Nature's building blocks. Oxford University Press, Oxford, 2011.

[9] J. Arblaster, Anal. Bioanal. Chem. 407 (2015) 3265–3265.

[10] J. Arblaster, Anal. Bioanal. Chem. 407 (2015) 6589–6590.

[11] C. Agte, H. Alterthum, Z. Tech. Phys. 11 (1930) 182–191.

[12] R.A. Andrievskii, N.S. Strel'nikov, N.I. Poltoratskii, E.D. Kharkardin, V.S. Smirnov, Powder Metall. Met. Ceram. 6 (1967) 65–67.

[13] O. Cedillos-Barraza, D. Manara, K. Boboridis, T. Watkins, S. Grasso, D.D. Jayaseelan, R.J.M. Konings, M.J. Reece, W.E. Lee, Sci. Rep. 6 (2016) 37962.

[14] E. Rudy, D.P. Harmon, United States Air Force Materials Laboratory Rept. AFML-TR-65-2, Part I, Vol. V, 1965.

[15] E. Rudy, United States Air Force Materials Laboratory Rept. AFML-TR-65-2, Part I, Vol. IV, 1965.

[16] E. Rudy, United States Air Force Materials Laboratory Rept. AFML-TR-65-2, Part II, Vol. I, 1965.

[17] A.I. Savvatimskiy, S.V. Onufriev, S.A. Muboyadzhyan, J. Eur. Ceram. Soc. 39 (2019) 907–914.

[18] Q.-J. Hong, A. van de Walle, Phy. Rev. B 92 (2015) 020104.

[19] J. Behler, M. Parrinello, Phys. Rev. Lett. 98 (2007) 146401.

[20] A.P. Bartók, M.C. Payne, R. Kondor, G. Csányi, Phys. Rev. Lett. 104 (2010) 136403.

[21] A.P. Thompson, L.P. Swiler, R.T. Christian, M.F. Stephen, J.T. Garritt, J. Comput. Phys. 285 (2015) 316–330.

[22] Kristof Schütt, Pieter-Jan Kindermans, Huziel Enoc Sauceda Felix, Stefan Chmiela, Alexandre Tkatchenko, and Klaus-Robert Müller. Schnet: A continuous-filter convolutional neural network for modeling quantum interactions. Advances in neural information processing systems, 30, 2017.

[23] L. Zhang, J. Han, H. Wang, Roberto Car, W. E, Phys. Rev. Lett. 120 (2018) 143001.

[24] H. Wang, L. Zhang, J. Han, W. E, Comput. Phys. Commun. 228 (2018) 178.

[25] L. Zhang, D.-Y. Lin, H. Wang, R. Car, W. E, Phys. Rev. Mater. 3 (2019) 023804.

[26] Y. Zhang, H. Wang, W. Chen, J. Zeng, L. Zhang, H. Wang, W. E, Comput. Phys. Commun. 253 (2020) 107206.

[27] G. Kresse, J. Furthmüller, Phys. Rev. B 54 (1996) 11169.

[28] G. Kresse, J. Furthmüller, Comp. Mater. Sci. 6 (1996) 15.

[29] J.P. Perdew, K. Burke, M. Ernzerhof, Phys. Rev. Lett. 77 (1996) 3865.

[30] P. E. Blöchl, Phys. Rev. B 50 (1994) 17953.

[31] G. Kresse, D. Joubert, Phys. Rev. B 59 (1999) 1758.

[32] H.J. Monkhorst, J.D. Pack, Phys. Rev. B 13 (1976) 5188.

[33] W. Jiang, Y. Zhang, L. Zhang, H. Wang, Chin. Phys. B 30 (2021) 050706.

[34] Y. Wang, L. Zhang, B. Xu, X.Y. Wang, H. Wang, Model. Simul. Mater. Sc. 30 (2021) 025003.

[35] T. Wen, R. Wang, L. Zhu, L. Zhang, H. Wang, D.J. Srolovitz, Z. Wu, npj Comput. Mater. 7 (2021) 206.

[36] F.-Z. Dai, B. Wen, Y. Sun, H. Xiang, Y. Zhou, J. Mater. Sci. Technol. 43 (2020) 168–174.

[37] F.-Z. Dai, Y. Sun, B. Wen, H. Xiang, Y. Zhou, J. Mater. Sci. Technol. 72 (2021) 8–15.

[38] F.-Z. Dai, B. Wen, Y. Sun, Y. Ren, H. Xiang, Y. Zhou, J. Mater. Sci. Technol. 123 (2022) 26–33.

[39] X. Wang, Y. Wang, L. Zhang, F.-Z. Dai, H. Wang, Nucl. Fusion 62 (2022) 126013.

[40] S.J. Plimpton, J. Comp. Phys. 117 (1995) 1-19.

[41] A.P. Thompson, H.M. Aktulga, R. Berger, D.S. Bolintineanu, W.M. Brown, P.S. Crozier, P.J. in't Veld, A. Kohlmeyer, S.G. Moore, T.D. Nguyen, R. Shan, M.J. Stevens, J. Tranchida, C. Trott, S.J. Plimpton, Comp. Phys. Comm. 271 (2022) 10817.

[42] F. Nogueira, Bayesian Optimization: Open source constrained global optimization tool for Python, https://github.com/fmfn/BayesianOptimization.

[43] M. Sheindlin, T. Falyakhov, S. Petukhov, G. Valyano, A. Vasin, Adv. Appl. Ceram. 117 (2018) s48–s55.

[44] A.I. Savvatimskiy, S.V. Onufriev, G.E. Valyano, S.A. Muboyadzhyan, J. Mater.

Sci. 55 (2020), 13559–13568.